\documentclass{aa}
\usepackage{graphicx}               
\usepackage{placeins}               
\graphicspath{{figures/}}           
\usepackage{txfonts}                
\usepackage{lscape}                 
\usepackage[
    colorlinks=true,
    urlcolor=cyan,
    linkcolor=blue,
    citecolor=blue
            ]{hyperref}

\usepackage{natbib}
\bibpunct{(}{)}{;}{a}{}{,}          

\usepackage{amsmath}
\usepackage{mathabx}
\usepackage{siunitx}
\DeclareSIUnit{\solarradius}{R_\sun}
\DeclareSIUnit{\solarmass}{M_\sun}
\DeclareSIUnit{\au}{au}
\DeclareSIUnit{\gauss}{G}
\DeclareSIUnit{\erg}{erg}
\DeclareSIUnit{\year}{yr}

\usepackage[normalem]{ulem}
\usepackage{color}

\begin{document}

\title{NIRwave: A wave-turbulence-driven solar wind model constrained by PSP observations}

\author{S. Schleich \inst{1}
                \and
        S. Boro Saikia \inst{1}
        \and
        U. Ziegler \inst{2}
        \and
        M. Güdel \inst{1}
        \and
        M. Bartel \inst{1}
}

\institute{University of Vienna, Department of Astrophysics, 
           Türkenschanzstrasse 17, 1180 Vienna, Austria\\
           \email{simon.schleich@univie.ac.at}
           \and
           Leibniz-Institut für Astrophysik Potsdam, 
           An der Sternwarte 16, 14482 Potsdam, Germany
}

\date{Received XXX; Accepted YYY}

 
\abstract
        {}
        {We generate a model description of the solar wind based on an explicit wave-turbulence-driven heating mechanism, and constrain our model with observational data.}
        {We included an explicit coronal heating source term in the general 3D magnetohydrodynamic code NIRVANA to simulate the properties of the solar wind. The adapted heating mechanism is based on the interaction and subsequent dissipation of counter-propagating Alfvén waves in the solar corona, accounting for a turbulent heating rate $Q_\mathrm{p}$. The solar magnetic field is assumed to be an axisymmetric dipole with a field strength of 1~G. Our model results are validated against observational data taken by the Parker Solar Probe (PSP).}
        {Our NIRwave solar wind model reconstructs the bimodal structure of the solar wind with slow and fast wind speeds of \SI{410}{\kilo\meter\per\second} and \SI{650}{\kilo\meter\per\second} respectively. The global mass-loss rate of our solar wind model is \SI{2.6e-14}{\solarmass\per\year}. Despite implementing simplified conditions to represent the solar magnetic field, the solar wind parameters characterising our steady-state solution are in reasonable agreement with previously established results and empirical constraints. The number density from our wind solution is in good agreement with the derived empirical constraints, with larger deviations for the radial velocity and temperature. In a comparison to a polytropic wind model generated with NIRVANA, we find that our NIRwave model is in better agreement with the observational constraints that we derive.}
        {}

%

%
%


\maketitle

\section{Introduction}
Among the many discoveries made during studies of the heliosphere over the last century is the ever-present mass flow radiating away from the Sun and throughout the Solar System \citep{biermann1951, parker1965}, now known as the `solar wind'. As the solar wind travels into the heliosphere, it carries angular momentum away from its source, slowing the rotation of the Sun \citep{kraft1967, weber1967}. This rotation rate is a major parameter influencing the solar magnetic field through a dynamo process \citep{charbonneau2020}, where the decrease in solar rotation leads to a decrease in magnetic activity \citep{skumanich1972, vidotto2014}. In addition to their association with this important aspect of stellar evolution \citep{cohen2014b, johnstone2015b, reville2016, pantolmos2017}, winds of low-mass main sequence stars like our Sun play an important role in (exo)planetary atmospheric evolution and escape \citep[e.g.][]{terada2009, lundin2011, kislyakova2014, blackman2018}. The impact stellar winds have on (exo)planetary atmospheres can be found in non-thermal erosion processes induced by high-energy particles making up the stellar mass flow \citep{lichtenegger2010, kislyakova2014}, as well as in their influence on exoplanetary magnetospheres \citep{cohen2014a, vidotto2014b}. Detailed investigations of the solar and stellar wind phenomena are therefore important for the characterisation of exoplanetary systems. 

Our proximity to the Sun enables us to study its wind properties in great detail using both observations and numerical modelling. The solar wind is driven in large part by thermal pressure gradients in the upper atmosphere of the Sun; however, the exact physical mechanisms that heat the solar corona and accelerate the solar wind are still not fully understood \citep{cranmer2009, cranmer2019, vidotto2021}. It is now recognised that the heating required for wind acceleration is driven by stellar magnetic fields. It is therefore imperative that we identify the connection between the rotation rate, magnetic field, and mass flow of the Sun. 

Our understanding of the physical properties associated with the solar wind have been greatly increased through observations from missions such as \emph{Ulysses} \citep{mccomas2003}, the Advanced Composition Explorer \citep[ACE,][]{stone1998}, and more recently the Parker Solar Probe \citep[PSP,][]{fox2016} and Solar Orbiter \citep{mueller2020}. We now know that the velocity distribution of the solar wind can be divided into two regimes: a `fast' wind and a `slow' wind. These are distinguished by median velocities of around \SI{750}{\kilo\meter\per\second} and \SI{400}{\kilo\meter\per\second}, respectively \citep{mccomas2008, johnstone2015a}. The fast solar wind originates from regions of open magnetic field lines extending into the heliosphere, called coronal holes, while the slow solar wind has its origin at the boundary region between open and closed field lines  \citep{krieger1973, cranmer2009}. The complexity of the distribution of the solar wind regimes varies with the solar cycle, where the clear separation during the solar minimum gives way to a much more complex structure close to the solar maximum \citep{mccomas2003}. The above {in situ} measurements provide necessary constraints to numerical models used to study both the solar wind and the winds of cool, solar-like stars by extension.

Since the detection of the first exoplanetary systems, thousands of new host stars have been identified, many of them similar to the Sun. It is therefore reasonable to make efforts in relating our increasing knowledge of the solar wind to other stellar hosts in an attempt to study the impact of stellar winds in exoplanetary systems. As the winds of low-mass, solar-like stars are very weak, investigations into their properties rely on indirect methods of their detection \citep{wood2021}. Comprehensive models of the solar wind are therefore necessary to study the characteristic parameters of these stellar winds \citep{cohen2014b, johnstone2015a, garaffo2015, alvarado-gomez2016b, reville2016, ofionnangain2019, shoda2019, BoroSaikia2020}. From the initial description of the solar wind, models describing the nature and behaviour of this mass flow have evolved to more accurately reflect {in situ} measurements of the corresponding parameters, and to include better physical descriptions, which take the driving forces of the solar wind into account. Extensive reviews of different approaches to model the solar wind and relate these models to stellar winds can be found, for instance, in \citet{gombosi2018} and \citet{vidotto2021}. 

One method commonly used to model the behaviour of the solar wind, and stellar winds of low-mass stars by extension, is to simplify the energy equation through the  introduction of a polytropic equation of state, taking the form of a power-law relation between the thermal pressure and the density 
$\left(p ~\propto~ \rho^{\Gamma}\right)$%
. Working with this relation implicitly assumes the heating of the expanding wind, where the temperature follows the mass density. The exponent $\Gamma$ therefore defines the temperature behaviour of such models. As polytropic wind models consider simplified assumptions about the physical processes behind the heating and acceleration of the wind, complexity can be added in other aspects: the polytropic index might be chosen to vary with respect to the distance from the central source \citep{johnstone2015a}, the domains of these models be can made larger in scale in order to study the global wind behaviour, and a more complex magnetic field geometry can be added \citep{vidotto2021, reville2016}.

A different way of generating solar wind models is to explicitly treat the energy deposition into the solar corona, which heats and drives the solar wind. The necessity for this arises from the fact that an expanding corona treated as a mono-atomic gas with an adiabatic index of $\Gamma = 5 / 3$ cannot produce an accelerated solar wind and therefore requires additional energy input to prevent adiabatic cooling from taking place \citep{cranmer2019}. Alfvén waves have long been proposed as a source of this coronal heating \citep{alfven1947}, where explicit solar wind models include the process of converting magnetic energy into thermal energy through the dissipation of turbulence arising from the interaction of counter-propagating waves \citep{cranmer2010, vanderHolst2014}.

A variety of numerical frameworks have been employed to reconstruct the properties of the solar wind and to investigate their generalisation to stellar winds. Examples of this include PLUTO \citep{mignone2007}, which was used to produce polytropic wind descriptions \citep[e.g.][]{reville2016, pantolmos2017, desai2020}, and the work of \citet{usmanov2000}, which introduced a multi-dimensional (2D) model of the solar wind that includes Alfvén wave turbulence, and has been further developed to a fully turbulence-driven, three-dimensional description \citep{usmanov2011, usmanov2014}. The Alfvén Wave Solar Model \citep[AWSoM,][]{vanderHolst2014}, as part of the Space Weather Modelling Framework \citep{toth2012}, was also used to generate wind descriptions based on the dissipation of Alfvén waves \citep[e.g.][]{oran2013, BoroSaikia2020}. Another model that incorporates the reflection of Alfvén waves is the work of \citet{shoda2019}. These models were refined to include many complex physical aspects of the solar wind, such as ion temperature anisotropy \citep{vanderHolst2014}, the parametric decay instability of Alfvén waves \citep{shoda2019}, and synoptic magnetograms as boundary conditions of the surface magnetic field \citep{oran2013, BoroSaikia2020}.

We do not aim to rival the complexity of these frameworks, but instead to produce a simple model description of the solar wind by coupling the 3D magnetohydrodynamic (MHD) code NIRVANA with a numerical framework for calculating the turbulent coronal heating rate \citep{cranmer2010}; this is the first time NIRVANA has been used in this manner. Observational data gathered from the Parker Solar Probe provide empirical constraints for the results of our simulations. Establishing a model description of the solar wind that is consistent with empirical data is an important step in studying the properties of solar-like stellar winds. This can be achieved by adjusting such a model to the input parameters of a larger sample of Sun-like exoplanet host stars.

This paper is structured as follows: In Sect.~\ref{sec:model_desc}, we describe the main parts of our model. Section~\ref{ssec:num_model} describes the code we used, as well as the initial and boundary conditions we implemented, and Sect.~\ref{ssec:wtd_source} summarises the wave-turbulence-driven heating routine that we coupled to the general MHD code. Section~\ref{sec:observ_constraints} is concerned with the evaluation of observational data to derive empirical constraints for our model. In Sect.~\ref{sec:results_disc}, we present the results of our solar wind model and discuss the limitations we encountered. We compare our model to the evaluated observational constraints in Sect.~\ref{ssec:nirwave_obs}, and to a polytropic wind model generated with NIRVANA in Sect.~\ref{ssec:nirwave_polytrope}. We summarise our findings in Sect.~\ref{sec:conclusion}.

\section{Model description}\label{sec:model_desc}
We used the generalised 3D MHD code NIRVANA \citep{ziegler2008} to create a model of the solar wind based on an explicit, wave-turbulence-driven heating mechanism \citep{cranmer2010}, as described below.

\subsection{Numerical model}\label{ssec:num_model}

NIRVANA is a grid-based \verb|C|-code that solves ---in its full complexity--- the time-dependent equations of a gas-dynamical system, including non-relativistic, compressible magnetohydrodynamics, dissipative processes, ambipolar diffusion, self-gravity, and source terms in force and energy \citep{ziegler2008}. It supports Cartesian, cylindrical, and spherical geometries in 2D and 3D, adaptive mesh refinement, and serial or parallel operation. 

For the application in this work, NIRVANA solves the non-dissipative equations of magnetohydrodynamics with source terms in force and energy: 
\begin{align}
        \frac{\partial \rho}{\partial t} + \nabla \left( \vec{m} \right) &= 0, \label{eq-cha4:mhd-continuity} \\
        \frac{\partial \vec{m}}{\partial t} + \nabla 
        \left[ 
        \vec{m} \vec{v} + p_{\mathrm{tot}} \cdot I - \frac{\vec{B}\vec{B}}{\mu} 
        \right] &= \rho \vec{f}_\mathrm{e} + \vec{f}_\mathrm{C}, \label{eq-cha4:mhd-momentum} \\
        \frac{\partial e}{\partial t} + \nabla 
        \left[ 
        \left( e + p_{\mathrm{tot}} \right) \vec{v} - \frac{\left( \vec{v} \vec{B} \right) \vec{B}}{\mu} 
        \right] &= \left( \rho \vec{f}_\mathrm{e} + \vec{f}_\mathrm{C} \right) \vec{v} + Q_\mathrm{p}, \label{eq-cha4:mhd-energy} \\
        \frac{\partial \vec{B}}{\partial t} - \nabla \times \left( \vec{v} \times \vec{B} \right) &= 0. \label{eq-cha4:mhd-induction}
\end{align}
Equations~(\ref{eq-cha4:mhd-continuity}) to (\ref{eq-cha4:total-energy}) represent the single-fluid MHD equations with additional source terms in force and energy, where \mbox{$\mu = \SI{4\pi d-7}{\volt\second\per\ampere\per\meter}$} represents the magnetic vacuum permeability, $p_{\mathrm{tot}}$ is the total pressure given by the sum of the gas pressure and the magnetic pressure, and $\vec{I}$ is the identity operator. There are eight primary variables in this set of equations: the fluid mass density $\rho$, the momentum density $\vec{m} = \rho \vec{v}$ (where $\vec{v}$ is the fluid velocity), the magnetic field $\vec{B,}$ and the total energy density $e$, which is the sum of the kinetic, thermal ($\varepsilon$), and magnetic energy density:

\begin{align}\label{eq-cha4:total-energy}
        e = \frac{\rho \vec{v}^{\, 2}}{2} + \varepsilon + \frac{\vec{B}^2}{2 \mu}.
\end{align}

The source terms are the gravitational force of the central body, which is represented by the mass-normalized, cell-centred force term $\vec{f}_\mathrm{e}$, and similarly the effects of a rotating frame of reference expressed through $\vec{f}_\mathrm{C}$, which is based on the average rotation rate of the Sun, taken to be $\omega = \SI{2.597d-6}{\radian\per\second}$. Additionally, $Q_\mathrm{p}$ represents the energy source term introduced by the wave-turbulence-driven heating mechanism more explicitly described in Section \ref{ssec:wtd_source}.

The simulation domain we use in this work is a three-dimensional spherical domain with coordinates ($r$, $\theta$, $\phi$), representing the distance from the origin of the domain, the polar angle, and the azimuth angle, respectively. Our domain has the dimensions  
$%
\left[\SI{1}{\solarradius}, \SI{40}{\solarradius} \right] \times \left[ 0, \pi \right] \times \left[ 0, 2\pi \right]%
$%
, with $\left[ 384, 64, 64 \right]$ grid cells, respectively, resulting in a uniform radial grid cell size of \SI{1.02e-1}{\solarradius}. NIRVANA currently only supports mesh refinement uniformly in all coordinate directions, and therefore we opted to increase the base number of radial grid cells instead of unnecessarily refining the angular coordinate directions as a compromise in order to decrease the computational load for these simulations.

We initialise the density with a nominal coronal value at the inner boundary and an ambient density with a drop of approximately five orders of magnitude throughout the rest of the domain. The solar magnetic field is initialised as a dipole, with the axis of the magnetic moment aligned with the rotational axis of the domain (representing the solar rotational axis) following the dipole equations for the field components in spherical geometry:
\begin{align}
        B_\mathrm{r} &= B_0 \cdot \left( \frac{r}{R_\Sun} \right)^{-3} \cdot \cos\theta, \\
        B_\theta &=  \frac{B_0}{2} \cdot \left( \frac{r}{R_\Sun} \right)^{-3} \cdot \sin\theta, \\
        B_\phi &= 0,
\end{align}
where $r$ denotes the radial distance from the centre of the domain and $\theta$ the polar angle. The parameter $B_0$ in these equations represents the polar dipole surface field strength. We selected a value of $B_0 = \SI{1}{\gauss}$, which is representative of the dipole surface field strength close to the solar minimum \citep{svalgaard2007}. The momentum density is initialised at a value of zero for all components and the total energy density is dynamically assigned in the first calculation step through our heating source term. The most significant initial parameters are listed in Table~\ref{table:moddesc_initialpar}.

We close the set of equations by relating the pressure and temperature of the system to the primary variables through an adiabatic equation of state:

\begin{align}
        \varepsilon &= \frac{p}{\gamma - 1}, \\
        T &= \frac{\bar{\mu} m_u p}{k_\mathrm{B} \rho}.
\end{align}

These equations relate the thermal energy density $\varepsilon$ ---which is calculated by rearranging Eq.~(\ref{eq-cha4:total-energy})--- to $p$ and $T$, where $\bar{\mu}$ represents the mean molecular weight, $m_u$ the atomic mass unit, and $k_\mathrm{B}$ the Boltzmann constant. We choose an adiabatic index $\gamma = 5/3$ and set $\bar{\mu} = 0.6$ to reflect a fully ionised hydrogen plasma with small amounts of heavier species \citep{desai2020}.

At the inner radial boundary, we fix the density, magnetic field strength, and momentum density values to the initial conditions, and the energy density is constrained through a zero-gradient condition. The outer radial boundary of our simulation domain is constrained by a simple outflow condition. We impose periodic boundary conditions at the azimuthal edges of the domain. Lastly, we constrain the polar edges of our domain through F-type boundary conditions (in NIRVANA, F-type boundary conditions are a type of natural boundary condition that allows free flow across the poles in full azimuthal domains) and employ the dual-energy treatment of NIRVANA, where the total and thermal energy equations are solved separately and synchronised in order to improve numerical robustness in regions of low plasma-$\beta$\footnote{For a more detailed description we reference the documentation of NIRVANA from  \url{https://gitlab.aip.de/ziegler/NIRVANA}.}.
\begin{table}
        \caption{Initial parameters of our simulation.}
        \label{table:moddesc_initialpar}
        \centering
        \renewcommand{\arraystretch}{1.25}
        \begin{tabular}{c c}
                \hline\hline
                Parameter & Value \\
                \hline
                $\rho_0$ & \SI{5e6}{\per\cubic\centi\meter} \\
                $B_0$ & \SI{1}{\gauss} \\
                $L_\perp \sqrt{B}$ & \SI{6d8}{\centi\meter\sqrt{\gauss}} \\
                $F/B_0$ &  \SI{2e4}{\erg\per\second\per\square\centi\meter\per\gauss} \\
                \hline
        \end{tabular}
        \tablefoot{These parameters represent input values of our model that strongly influence our steady-state solar wind. $\rho_0$ represents the density value at the radial base of our simulation domain, $B_0$ the polar surface magnetic field strength we have chosen, and $L_\perp$ and $F/B_0$ are the free parameters governing the output of the wave-turbulence-driven heating routine.}
\end{table}

We search for a steady-state solution of our simulation by evaluating the incremental change in the plasma parameters $\rho$, $T,$ and $v_\mathrm{r}$ over large differences in domain time. For this, we compare the plasma parameters between simulation states separated by \SI{5e4}{} time steps and accept a steady state once the relative changes between these two snapshots fall below a sufficiently small threshold.

\subsection{WTD heating source term}\label{ssec:wtd_source}

Our model couples NIRVANA to a wave-turbulence-driven heating routine described by \citet{cranmer2010}. It is based on the reflection and subsequent dissipation of Alfvén waves travelling as perturbations along magnetic flux tubes. The computation of the local turbulent heating rate $Q_\mathrm{p}$ as an energy source depends on the fluid density $\rho$, fluid velocity $\vec{v}$, magnetic field strength $B,$ and additionally on the distance from the centre of the domain $r$. The output of the heating routine is additionally modified by the chosen correlation length scale ($L_\perp$) and wave energy flux parameter ($F/B_0$) as free parameters. A full derivation of the analytic background can be found in \citet{cranmer2010}; we provide an outline of the computational routine below.

The heating rate is characterised through the interaction and dissipation of outward propagating waves and reflected inward propagating waves. These are characterised through the Elsässer components $z_-$ and $z_+$, respectively. They define a reflection coefficient $\mathcal{R}$, which compares the amplitudes of the two waves and characterises the reflection `efficiency':
\begin{equation}\label{eq:reflection-coefficient}
    \mathcal{R} = \frac{\vert z_+ \vert}{\vert z_- \vert},
\end{equation} 
with the magnitude of outward-propagating waves always being larger than the magnitude of inwardly propagating waves, or \mbox{$\mathcal{R} < 1$}.

The reflection coefficient is crucial in calculating the heating rate, as it defines the amplitude of the reflected wave. $\mathcal{R}$ depends on the Alfvén wave frequency $\omega_i$, and \citet{cranmer2010} outlines a routine to calculate a radius- and frequency-dependent $\mathcal{R}$ based on the limiting cases of very low ($\mathcal{R}_\text{zero}$) and very high ($\mathcal{R}_\infty$) frequencies. The turbulent heating rate $Q_\mathrm{p}$ is then calculated through the spectrum-weighted Elsässer variables $Z_\pm$ following the equation
\begin{equation}\label{eq-cha4:turb-heating}
    Q_\mathrm{p} = \rho \cdot \frac{Z_+ Z_- (Z_+ + Z_-)}{4 L _\perp}.
\end{equation}

The necessary Elsässer amplitudes are calculated through
\begin{equation}\label{eq-cha4:spectrum-w-els}
    Z_- = \sqrt{\frac{4 U_\mathrm{A}}{\rho \left(1 + \left< \mathcal{R} \right>^2\right)}},
\end{equation}
and related by the spectrum-weighted reflection coefficient following Eq.~(\ref{eq:reflection-coefficient}). This coefficient $\left< \mathcal{R} \right> = \sum_i \mathcal{R}(\omega_i) \cdot f_i$ is calculated through a discretization into 17 frequency bins spanning almost five orders of magnitude, with weights $f_i$ corresponding to a high-frequency-dominated power spectrum for Alfvén waves \citep{cranmer2005}.

The output of Eq.~(\ref{eq-cha4:turb-heating}) is additionally dependent on two free parameters. The transverse correlation length scale $L_\perp$ is a measure for the correlation length of the largest turbulent eddies and follows the scaling relation described by \citet{hollweg1986}:
\begin{align}
    L_\perp \, \propto \, B^{-1/2},
\end{align}
where we find that a scaling factor of \SI{6d8}{\centi\meter\sqrt{\gauss}} best reproduces our solar wind constraints derived from observational data, being an intermediate value between the factors of \SI{11.55d8}{} and \SI{2.876d8}{} based on normalisation from previous results (see \citealt{cranmer2010} and references therein). 

The other free parameter of the calculation routine is the wave energy flux $F$ per unit magnetic field strength $B_0$. For dispersionless Alfvén waves, the conservation of wave action implies a constant value of the parameter $F/B_0$, given by the expression
\begin{align}
    \frac{F}{B_0} = \frac{\left(v + v_\mathrm{A}\right)^2 U_\mathrm{A}}{v_\mathrm{A} B_0} = \mathrm{ const.},
\end{align}
where $v_\mathrm{A} = B / \left(\mu_0 \rho \right)^{1/2}$ denotes the Alfvén velocity and $U_\mathrm{A}$ the wave energy density \citep{jacques1977}. This allows the computation of $U_\mathrm{A}$ necessary for Eq.~(\ref{eq-cha4:spectrum-w-els}). We find that, within the range previously described by \citet{cranmer2010}, a value of $F/B_0 = \SI{2.0e4}{\erg\per\second\per\square\centi\meter\per\gauss}$ best reproduces the bimodal solar wind distribution.

We coupled the calculation routine outlined above to NIRVANA to include it as an energy source term (represented in Eq.~(\ref{eq-cha4:mhd-energy}) by the parameter $Q_\mathrm{p}$) to drive the solar wind. Natively, NIRVANA supports user-defined heating (and cooling) functions with dependencies on fluid mass density $\rho$ and temperature $T$. In order to facilitate the inclusion of a more complex heating source term, we modified the code to allow a user-defined heating function depending on the parameter space described above. A summary of our initial conditions is displayed in Table~\ref{table:moddesc_initialpar}.

\section{Observational data}\label{sec:observ_constraints}
To provide empirical constraints for our numerical model, we evaluate observational data from the Solar Probe Cup \citep[SPC,][]{case2020} and Solar Probe ANalyzer -- Ions \citep[SPAN-I,][]{livi2022} instruments of the Solar Wind Electrons Alphas and Protons investigation \citep[SWEAP,][]{kasper2016} carried by the Parker Solar Probe \citep[PSP,][]{fox2016}. The SPC operates through the selection of particles based on their energy-to-charge ratio $\left( E/q \right)$ by a modulating high-voltage (HV) grid. Non-repelled particles induce a signal on a collector plate, which makes up the raw data acquired from SPC (we refer to \citealt{case2020} for a detailed description of the instrument). The SPAN-Ion instrument is an electrostatic analyser, and part of the SPAN-A module pointed in the ram direction of the spacecraft to analyse the three-dimensional distribution function of solar wind ions. Ions are selected by elevation angle and then energy-to-charge ratio as they pass through the electrostatic analyser. An azimuthal distribution is resolved with a dedicated anode board (see \citealt{livi2022} for a more detailed description of the instrument). 

In general, the orbit of PSP is divided into two regimes: `cruise' (at heliocentric distances $>\SI{54}{\solarradius}$) and `encounter' (at heliocentric distances $<\SI{54}{\solarradius}$). During the cruise phase of an orbit, both instruments operate more sporadically, while the measurement cadence is increased significantly during the encounter phase, when the spacecraft is in a region of increased scientific interest \citep{case2020, livi2022}. SPC and SPAN-I are complementary to each other: SPC is designed to measure the ion particle flux in the outer phase of an encounter, where the solar wind flows are primarily radial, while SPAN-I is optimised to measure these flows close to the perihelion, where they might be strongly non-radial and out of view for SPC \citep{livi2022}.

Data processing reduces the raw information sent by the instruments to Level-2 and Level-3 science data products. SPC Level-3 data products represent solar wind plasma properties calculated both through fits of Maxwellian distributions to the current spectra, as well as through moment calculations of a reduced-velocity distribution function (RDF) gained from the differential energy flux \citep{case2020}. These properties include the proton population density, velocity vector components, and thermal speed ($w = \sqrt{2 k_\mathrm{B} T / m}$), from which we derive the scalar temperature. Similarly, Level-3 SPAN-I data products for the proton population contain the density, velocity vector components, and scalar temperature as (partial) moments of the plasma velocity distribution \citep{livi2022}.

Both instruments supply their measurements for the velocity, density, and temperature with respect to both the spacecraft frame (SC) and an inertial radial-tangential-normal frame (RTN) in heliocentric inertial coordinates (HICs). The latter are in reference to the solar equatorial plane and can therefore be used to place the measurements both with respect to heliocentric distance and heliolatitude. For SPC, the velocities reported in this frame have already been corrected for the variable motion of the spacecraft, depending on the distance. We employ the same correction for reported SPAN-I velocities with the provided ancillary data. We evaluate the Level-3 data products derived from observations during the designated encounter phases 7, 8, and 9 to determine empirical constraints for our solar wind model, making use of the parameters provided through the moment calculations. The measurement data are publicly released in regular intervals\footnote{\url{http://sweap.cfa.harvard.edu/pub/data/sci/sweap/}}, and details for these measurement periods are listed in Table~\ref{table:encounter_phase_info}.
\begin{table*}[]
    \centering
    \renewcommand{\arraystretch}{1.2}
    \caption{Evaluated PSP encounter phases.}
    \label{table:encounter_phase_info}
    \begin{tabular}{c c c c c c}
    \hline\hline
        Encounter & Start & End & Perihelion & \multicolumn{2}{c}{Number of measurements} \\
          & Date & Date & $\left[\si{\solarradius}\right]$ & SPC & SPAN-I \\
        \hline
        7 & 12 Jan 2021 & 23 Jan 2021 & 20.3 & \SI{1.580e5}{} & \SI{4.048e4}{} \\
        8 & 24 Apr 2021 & 4 May 2021 & 15.9 & \SI{2.471e5}{} & \SI{6.074e4}{} \\
        9 & 4 Aug 2021 & 15 Aug 2021 & 15.9 & \SI{1.733e5}{} & \SI{7.640e4}{} \\
    \hline
    \end{tabular}
    \tablefoot{PSP observational periods for which we have evaluated the provided Level-3 data products. The number of measurements for each encounter is calculated after we have evaluated the data according to our quality restrictions, but before determining the ten-second-average values.}
\end{table*}

For SPC, we derive the proton population number density, radial velocity component, and temperature from the available distribution moments, and reduce the measurement data conservatively by only accepting data points with a \verb|GENERAL_FLAG| of 0, indicating that the measurement has been made under ideal conditions. For SPAN-I, we derive the proton population density, radial velocity component, and temperature from the available (partial) moments of the plasma velocity distribution. We make sure that the solar wind is in the field of view (FOV) of the instrument by only accepting data points where the azimuthal flux peak is at or below $\phi=\SI{150}{\degr}$ \citep{livi2022}. To evaluate the measurements of SPC and SPAN-I together, we determine the ten-second-average of each parameter for both instruments to account for different measurement cadences.

\begin{figure}
    \centering
    \includegraphics[width=\hsize]{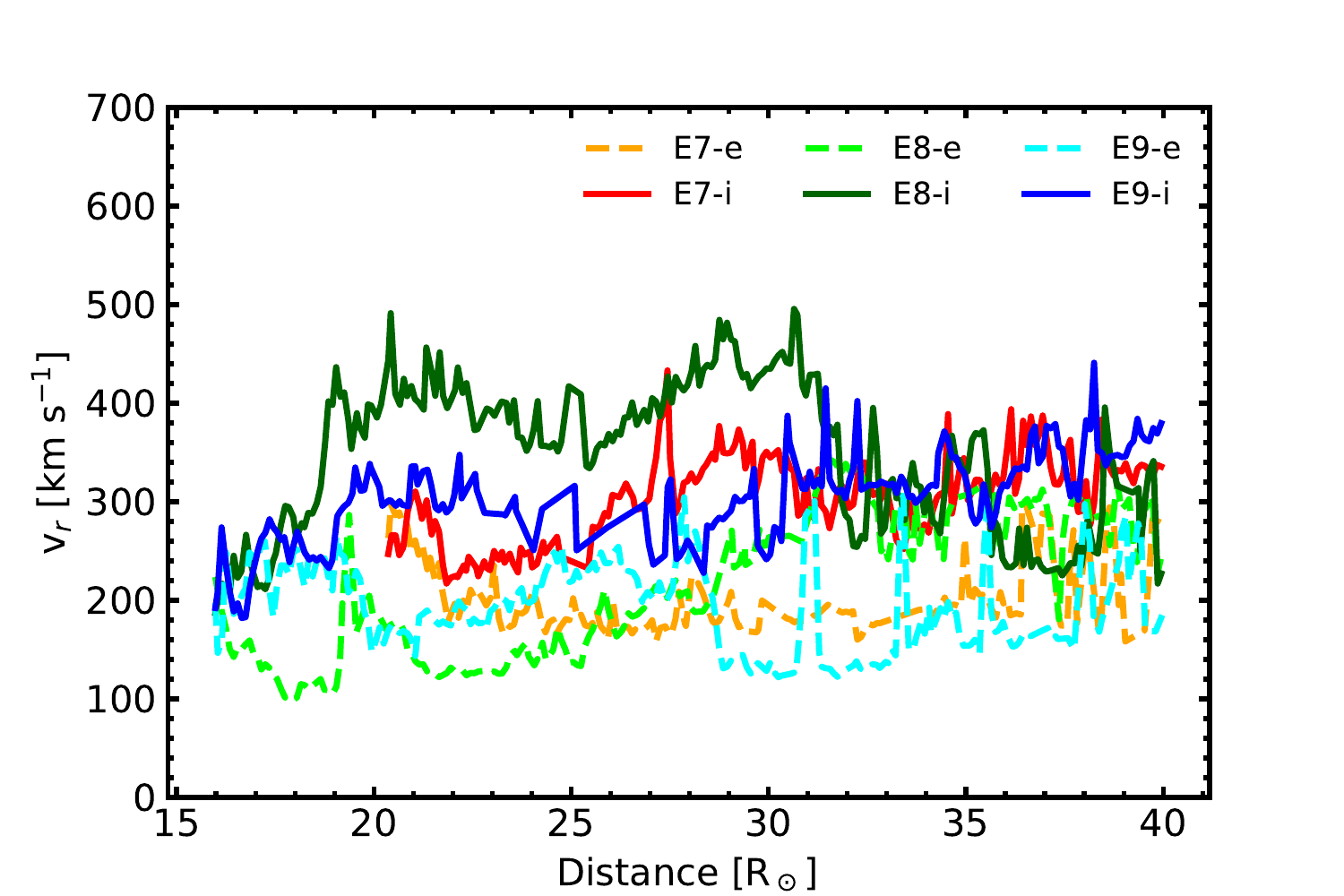}
    \includegraphics[width=\hsize]{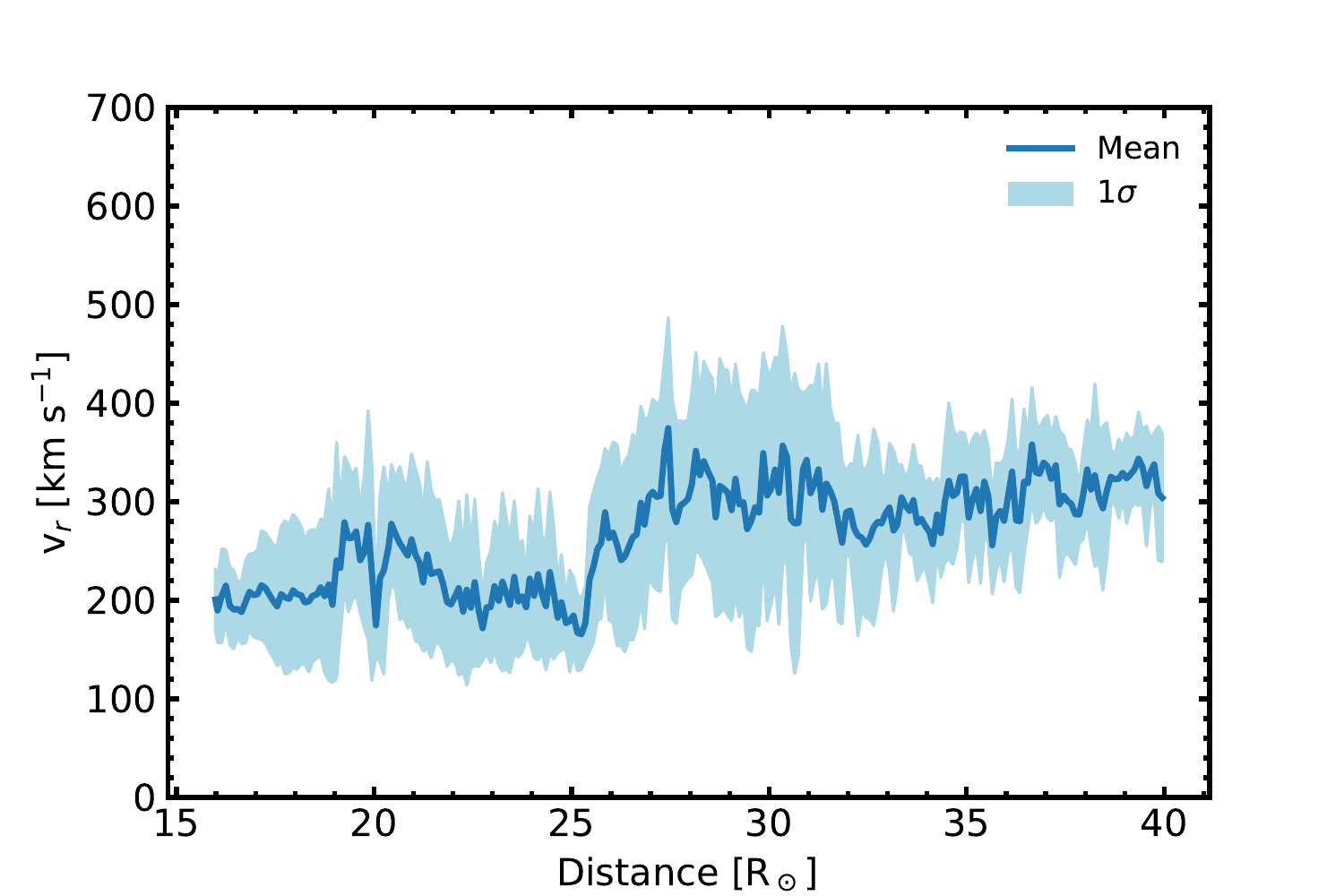}
    \caption{Ten-second-averaged radial-velocity measurements of SPC and SPAN-I with respect to heliocentric distance.
    Upper panel: Measurement data binned in intervals of \SI{0.1}{\solarradius} for the encounter periods 7, 8, and 9 of PSP. The displayed data are split into ingress (i) and egress (e) phases for each period.
    Lower panel: Mean and 1$\sigma$ range of these measurements.}
    \label{fig:obs_radvel}
\end{figure}

We further evaluate the measured parameters by binning the data in intervals of \SI{0.1}{\solarradius} to coincide with the radial cell size of our simulation grid and determining the mean and standard deviation for each parameter in each bin. For the combined encounter periods, the range of heliocentric distances covered after our reduction of the measurement data extends between \SI{40}{\solarradius} (the outer edge of our simulation domain) and \SI{15.9}{\solarradius}. In total, we included approximately \SI{7.56e5}{} measurements across this range to ascertain empirical constraints for our numerical model. Figure~\ref{fig:obs_radvel} shows the radial-velocity measurements taken by SPC and SPAN-I that we evaluated, as well as the mean values we derived as empirical constraints for this parameter. Figure~\ref{fig_app:spc_ingress_egress_obs} additionally displays the corresponding number density and temperature measurements and evaluations. The combination of both instruments provides an overview of the general profiles of the solar wind parameters during the evaluated encounter periods. The complementary nature of SPC and SPAN-I is illustrated in Figs.~\ref{fig:data_per_bin} and \ref{fig_app:psp_8i_obs}. The majority of measurements at distances $> \SI{25}{\solarradius}$ come from SPC, whereas SPAN-I provides most of the data points at distances $<\SI{25}{\solarradius}$.  

We note that we have omitted the observations taken during the designated encounter period 10. The SPC instrument turned off on 18 Nov 2022, when a safety limit was exceeded during high-speed solar wind stream observations, and did not take further measurements during this encounter\footnote{The anomalies recognised during every encounter period are published at \url{http://sweap.cfa.harvard.edu/Data.html} in data release notes for both SPC and SPAN-I}. The data provided from this encounter phase stop at a radial distance of approximately \SI{40}{\solarradius} and do not fall within our simulation domain. We also compare the results of our simulation to previous analyses of measurements by the FIELDS suite of instruments \citep{bale2016} as part of the PSP mission (we refer to \citealt{telloni2021, telloni2022}, further discussed in Sect.~\ref{sec:results_disc}).

\section{Results and discussion}\label{sec:results_disc}
We carried out steady-state solar wind simulations with the initial parameters listed in Table~\ref{table:moddesc_initialpar}. From the steady-state solution of our model, we determine the radial velocity ($v_\mathrm{r}$), number density ($n_\mathrm{p}$), and temperature ($T$) within the domain. The evaluation of observational data taken by PSP has allowed us to put empirical constraints on these parameters. We note that for the comparison between the results of our simulations and the observational data from PSP, we only display the evaluated solar wind measurements up to a distance of \SI{40}{\solarradius}, which coincides with the outer boundary of our domain.

\begin{figure*}
    \centering
    \includegraphics[width=\hsize]{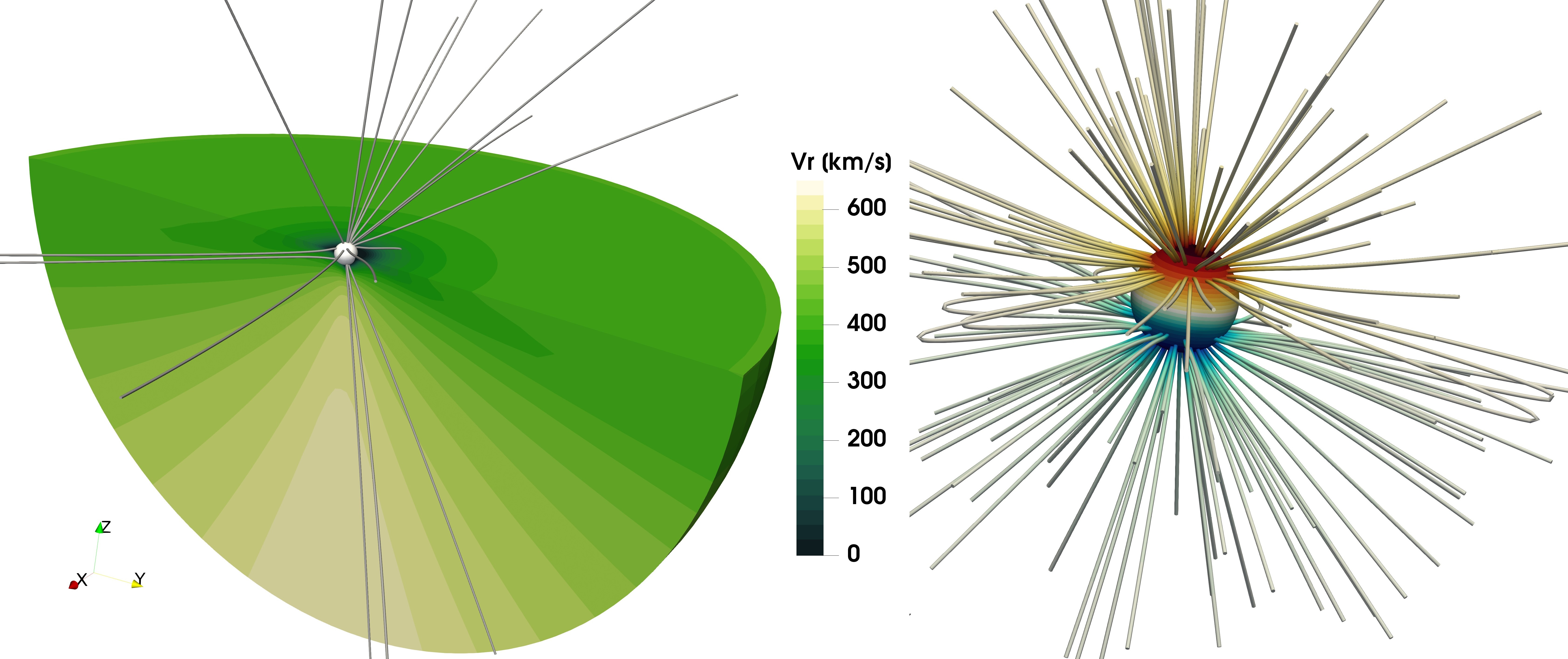}
    \caption{Three-dimensional structure of our steady-state solar wind simulation. Left: Radial-velocity structure (colour map) and magnetic field lines (grey lines). Right: Close-in structure of the magnetic field within \SI{10}{\solarradius}. The spherical surface denotes the inner boundary of our domain and shows the surface field strength of the radial magnetic field component $B_\mathrm{r}$ as a colour gradient between the north and south pole of the domain. The grey lines represent the connected field lines.}
    \label{fig:3d-solarwind}
\end{figure*}

\begin{figure}
    \centering
    \includegraphics[width=\hsize]{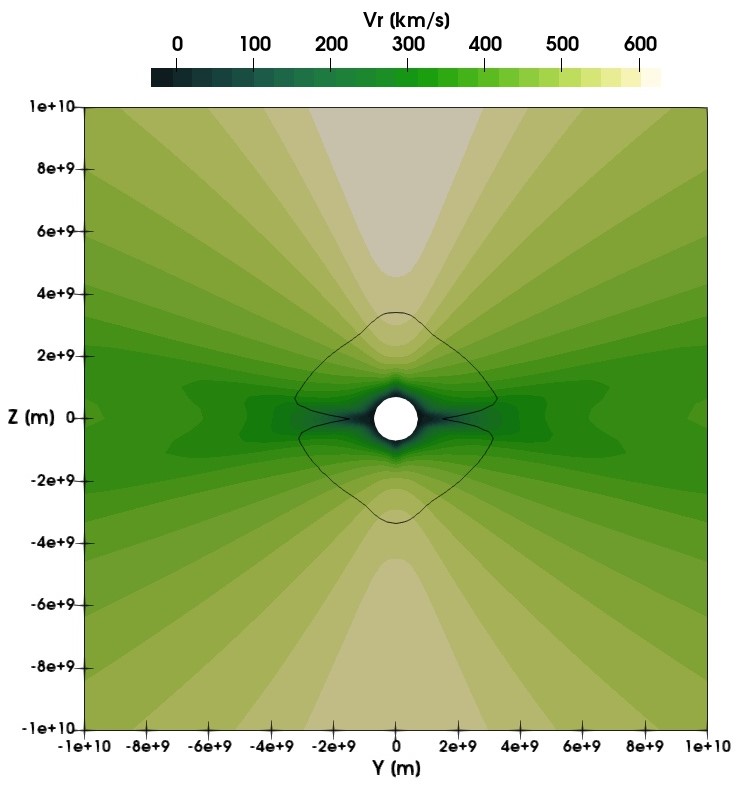}
    \caption{Meridional slice of the simulation domain (X = 0) extending to $\pm\SI{15}{\solarradius}$ in both Y and Z directions. The colour map displays radial-velocity values, and the solid black outline denotes the location of the Alfvén surface.}
    \label{fig:2d_solarwind}
\end{figure}

We show the three-dimensional structure of our steady-state simulation in Fig.~\ref{fig:3d-solarwind}, displaying the meridional and equatorial distribution of $v_\mathrm{r}$ and global magnetic field structure (left panel), which takes on a predominantly radial structure characterised by $\vert B_\mathrm{r} / B \vert > 0.95$ beyond approximately \SI{2}{\solarradius}, which is comparable to the canonically assumed distance of the source surface in potential field source surface (PFSS) models \citep{riley2006}. The magnetic field structure close to the inner boundary of the simulation domain is shown in Fig.~\ref{fig:3d-solarwind} (right panel).  Our simulation successfully reproduces the characteristic bimodal structure of the solar wind. The azimuthal distribution of the solar wind is highly symmetrical within the equatorial plane, while the radial-velocity structure within the meridional plane shows a strong variation correlated to heliolatitude. We display a meridional slice of the radial velocity in Fig.~\ref{fig:2d_solarwind}, illustrating the separation into two wind regimes connected to the polar angle. We compare this to the results from, for instance, \citet{sokolov2013}, which show an acceleration of the fast wind to \SI{400}{\kilo\meter\per\second} within \SI{4}{\solarradius},  or \citet{johnstone2015a}, where the majority of the fast-wind acceleration takes place within approximately \SI{10}{\solarradius}. Our results show a comparable, but more rapid acceleration of the solar wind. We therefore find that the transition from sub-Alfvénic to the super-Alfvénic velocity regimes at the distance of the Alfvén surface, where $v_\mathrm{A} / v_\mathrm{r} = 1$ (denoted by the black outline in Fig.~\ref{fig:2d_solarwind}), sits close the centre of our domain at a distance of between 3 and \SI{6}{\solarradius}. This is shorter than what is predicted by other models \citep[see][and references therein]{chhiber2019}, and is also shorter than the distance
derived from measurements by PSP, which has crossed the Alfvén critical surface during several intervals at distances of between 15 and \SI{20}{\solarradius} \citep{kasper2021}. The Alfvén surface from our model underestimates these distances by approximately a factor of 3. As the location of the Alfvén surface is also influenced by the magnetic field configuration through the definition of the Alfvén velocity $v_\mathrm{A}$, a discrepancy between our model and previous measurements and models in its size coincides with our choice of a simplified dipole to represent the solar magnetic field. Additionally, the location of the Alfvén surface depends on the acceleration profile of the wind solution. The more rapid acceleration of the solar wind close to the inner boundary in our result compared to results from other simulations also contributes to the smaller size of the Alfvén surface.

\subsection{Comparison to PSP observations}\label{ssec:nirwave_obs}

\begin{figure}
    \centering
    \includegraphics[width=\hsize]{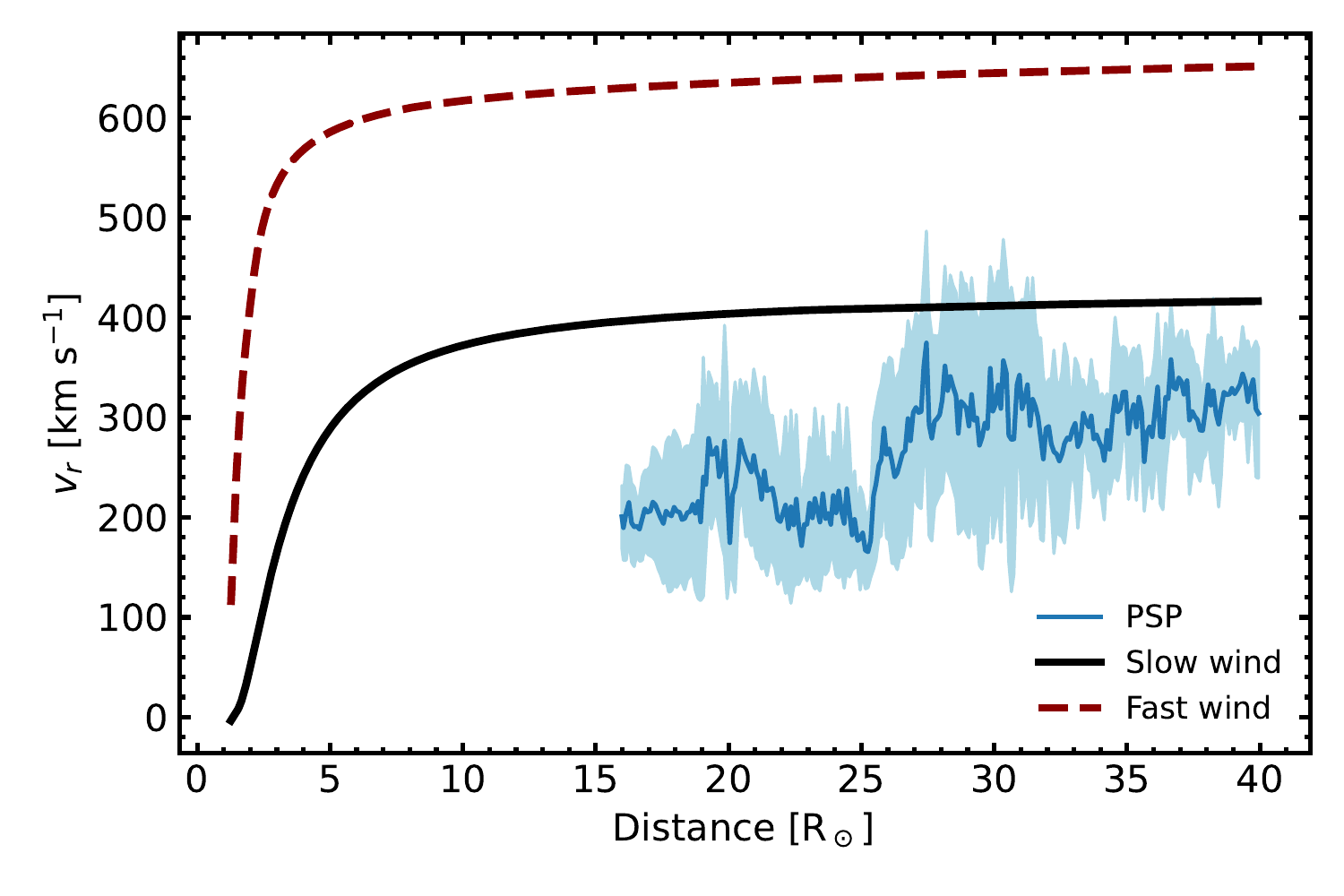}
    \caption{Comparison between the radial-velocity profiles of our solar wind simulation and the evaluated measurements from PSP. The solid black line represents a radial profile of $v_\mathrm{r}$ taken from the equatorial plane of the simulation domain (representing the slow solar wind), the red dashed line a polar radial profile (representing the fast solar wind), and the blue solid line represents the mean $v_\mathrm{r}$ measurements from PSP with a variation of 1$\sigma$.}
    \label{fig:vr_profiles}
\end{figure}

To compare the results from our simulations with the observational data from PSP, we produced radial profiles of the major parameters from both the equatorial and polar region of the simulation domain. As the trajectory of PSP is characterised by small heliolatitudes during the encounter phases, we expect to reproduce the empirical constraints with the equatorial profile of our simulation results. Figure~\ref{fig:vr_profiles} displays the profiles and observational constraints for the radial velocity in \si{\kilo\meter\per\second}. We see that the equatorial profile of our simulation is centred at approximately \SI{410}{\kilo\meter\per\second} and overestimates the empirical constraints within a factor of approximately 1.5, where the discrepancy between the constraints and simulation result is more pronounced at smaller heliocentric distances. The corresponding polar profile from our simulation result reaches a radial velocity of approximately \SI{650}{\kilo\meter\per\second}. We therefore see the characteristic bimodal structure of the solar wind velocity as described in, for instance, \citet{mccomas2003}. Under the assumption that the solar wind speed reaches a terminal value after a period of heavy acceleration close to the Sun, the radial velocity in our fast wind regime is lower than that suggested by the work of \citet{BoroSaikia2020} and that suggested by the fast wind measurements of approximately \SI{760}{\kilo\meter\per\second} taken by \emph{Ulysses} \citep{mccomas2000, mccomas2003}. The bimodality of the solar wind emerges from Eq.~(\ref{eq-cha4:turb-heating}), which depends on the magnetic field strength throughout the domain introduced by the dipole field and leads to a non-uniform distribution of the turbulent heating rate $Q_\mathrm{p}$ with respect to the polar angle, facilitating the emergence of the characteristic two velocity regimes of the solar wind with respect to heliolatitude.

\begin{figure}
    \centering
    \includegraphics[width=\hsize]{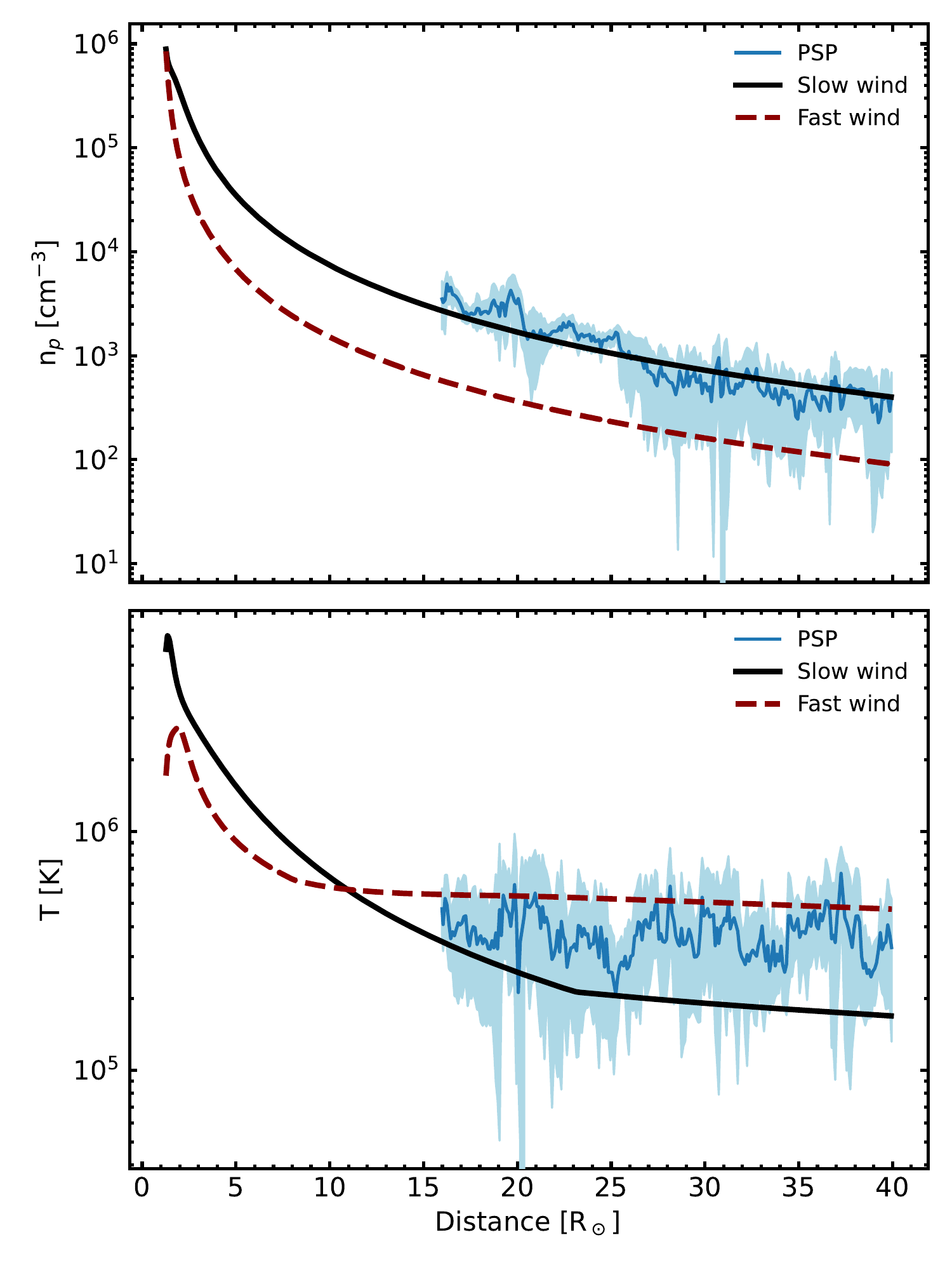}
    \caption{Same as Fig.~\ref{fig:vr_profiles}, but for the number density ($n_\mathrm{p}$, upper panel) and temperature ($T$, lower panel) of the simulation and observations.}
    \label{fig:npT_profiles}
\end{figure}

Figure~\ref{fig:npT_profiles}  displays the radial profiles for the number density ($n_\mathrm{p}$) in units of \si{\per\cubic\centi\meter} and temperature ($T$) in \si{\kelvin}. The equatorial profile of $n_\mathrm{p}$ from our simulation coincides very closely with the empirical constraints. We see a similar radial behaviour for the number-density profiles of the slow and fast wind regimes of our simulation, where in the case of the fast wind, $n_\mathrm{p}$ is lower by approximately a factor of 4. For the temperature, we see a pronounced difference in the radial behaviour of the fast- and slow-wind-domain profiles. The observed temperature is an intermediate between the two wind regimes from the simulation, where the discrepancy becomes larger for increasing heliocentric distances, but is confined to approximately a factor of 1.5 for both wind regimes.

We also derive the mass-loss rate $\dot{M}$ from our simulation parameters. The mass loss induced by the solar wind is characterised through a surface integral of the radial momentum density,
\begin{align}\label{eq:mass_loss_int}
    \dot{M} = \oint_S \rho v_\mathrm{r} \, dS.
\end{align}

\begin{figure}
    \centering
    \includegraphics[width=\hsize]{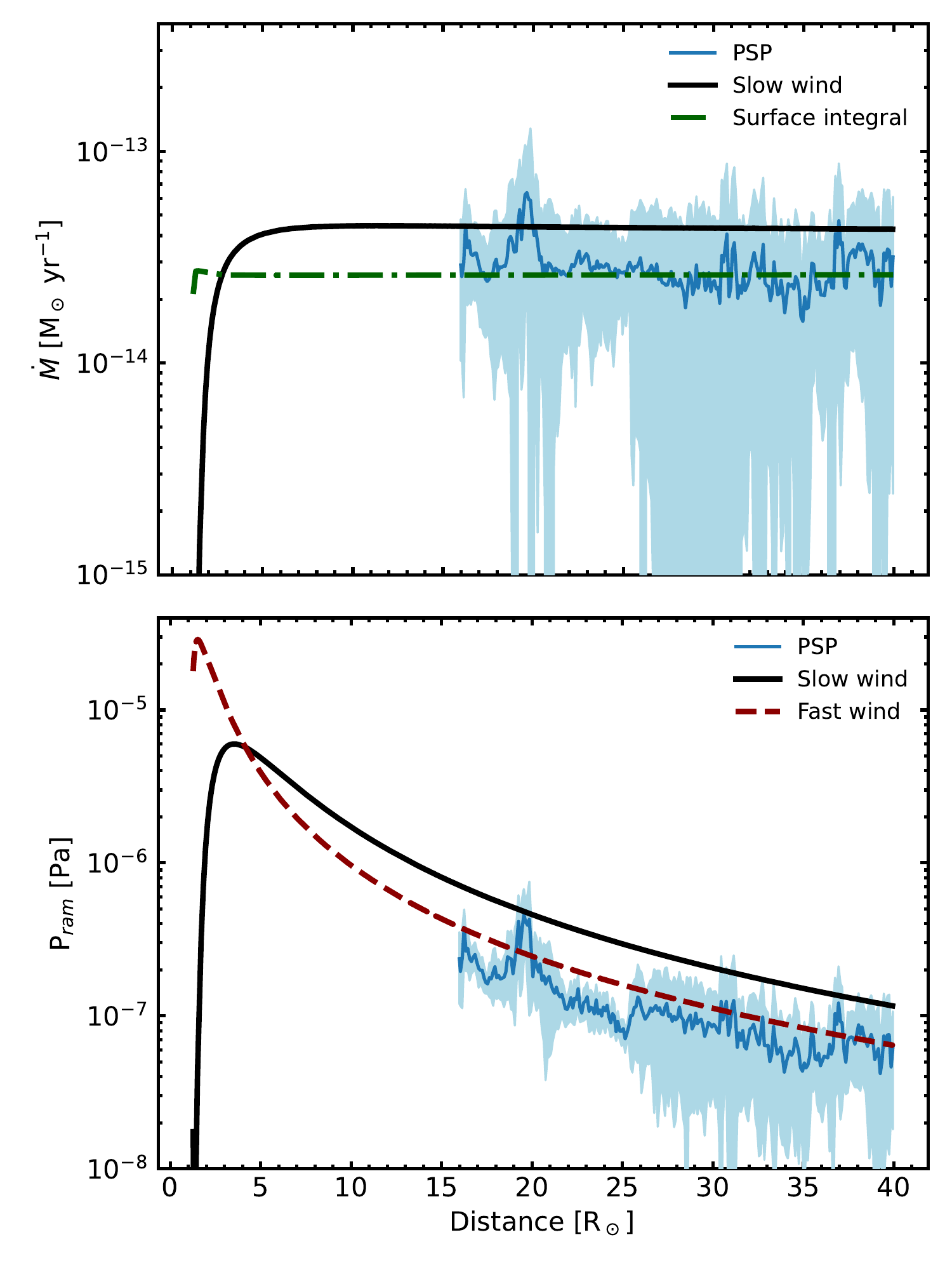}
    \caption{Mass-loss rate $\dot{M}$ and ram pressure $P_\mathrm{ram}$ as radial profiles compared to the empirical constraints derived from observations taken by PSP. In both panels, the black solid line represents an equatorial radial profile taken from the simulation and the blue line and filled-in area represent the mean observational values from PSP and a deviation of 1$\sigma$, respectively. 
    Upper panel: Mass-loss rate $\dot{M}$. The green dashed-dotted line represents the mass-loss rate derived from surface integrals following Eq.~(\ref{eq:mass_loss_int}) for radial shells distributed throughout the simulation domain. 
    Lower panel: Ram pressure $P_\mathrm{ram}$. The red solid line represent the radial polar profile from our simulation results.}
    \label{fig:massloss_rampressure_comp}
\end{figure}

Except for close distances to the Sun, where the solar magnetic field is not dominated by its radial component (i.e. where not all field lines are open), the solar mass-loss rate is constant with respect to radial distance \citep{cohen2011}, and we display the corresponding radial profiles in Fig.~\ref{fig:massloss_rampressure_comp} in units of \si{\solarmass\per\year}. As PSP orbits close to the equatorial plane of the Sun, we extrapolate a global mass-loss rate from the observational data through $\dot{M} = 4 \pi r^2 \rho v_\mathrm{r}$, where $r$ is any given distance from the Sun, $\rho$ is the measured mass density, and $v_\mathrm{r}$ the measured radial velocity. We compare this to the equatorial profile from our steady-state solutions, where the mass-loss rate is \mbox{$\dot{M} = \SI{4.3d-14}{\solarmass\per\year}$} (solid black line in the top panel of Fig.~\ref{fig:massloss_rampressure_comp}). Additionally, we evaluate the mass-loss rate from our simulation results through Eq.~(\ref{eq:mass_loss_int}) by taking the integrated values of the momentum density from radial shells distributed throughout the simulation domain, resulting in a value of approximately $\dot{M} = \SI{2.6e-14}{\solarmass\per\year}$ (dashed-dotted green line in the top panel of Fig.~\ref{fig:massloss_rampressure_comp}). Both values are in good agreement with our empirical constraints. solar wind simulations of the solar-cycle minimum by \citet{alvarado-gomez2016b} and \citet{BoroSaikia2020} also agree with our results.

We additionally evaluate the ram pressure $P_\mathrm{ram}$ of the solar wind, as it is a major influencing factor in the shape of planetary magnetospheres. The calculation follows the equation below,
\begin{align}\label{eq:rampressure}
    P_{\mathrm{ram}} = \rho v_\mathrm{r}^2,
\end{align}
where $\rho$ is the mass density and $v_\mathrm{r}$ the radial-velocity component. The bottom panel of Fig.~\ref{fig:massloss_rampressure_comp} shows the radial equatorial and polar ram-pressure profiles from our simulations in comparison to the ram-pressure values derived from observations taken by SPC and SPAN-I in units of \si{\pascal}. The slow wind displays a ram-pressure value of \SI{1.2d-7}{\pascal} at \SI{40}{\solarradius}, compared to the observational value of \SI{6.3d-8}{\pascal}. The fast wind exhibits a ram pressure of \SI{6.4d-8}{\pascal}. Both the equatorial and polar profile show similar values, as the differences in $\rho$ and $v_\mathrm{r}$ between the slow and fast wind regime counteract one another in Eq.~(\ref{eq:rampressure}). The ram-pressure profile derived from the measurement data is in better agreement with the fast-wind profile from our simulation, but both cases are in agreement within a factor of 3 with the empirical constraints and follow the same radial gradient as the observational data indicate. We note that the deviations of $\rho$ and $v_\mathrm{r}$ between our simulation results and the observational constraints propagate into the evaluation of $\dot{M}$ and $P_{\mathrm{ram}}$, as these are derived quantities of the primary parameters.

We find that the results of our NIRwave simulation are in reasonable agreement with the observational constraints we have evaluated, but the displayed parameter values and behaviours do not exactly coincide with previously established results and observations, with some being in better agreement than others. The number density measured by PSP closely agrees with the results from our simulation corresponding to the slow wind profile, while the radial velocity and temperature associated with the slow wind regime for our simulation deviate from the derived empirical constraints within a factor of approximately 1.5. The radial velocity for the fast wind from our simulations underestimates observations by \emph{Ulysses} \citep{mccomas2000, mccomas2003} by approximately 15\%, and the majority of our wind acceleration happens close to the centre of the domain within \SI{5}{\solarradius}. This subsequently leads to a radially small Alfvén surface below \SI{6}{\solarradius}. The difference in solar wind parameter measurements from PSP and our simulation results can be attributed to several factors. We implemented a simplified dipole description of the solar magnetic field, which reproduces the overall bimodal structure of the solar wind distribution but does not account for a more detailed description of the magnetic field structure at the inner boundary of the simulation domain. Other studies used synoptic magnetograms to generate initial conditions for this parameter \citep[e.g.][]{reville2020, chhiber2021, vanderholst2022}. Due to the uniform grid decomposition used in our simulations, we also cannot achieve the necessary resolution to treat the transition region within the domain without unnecessarily refining regions further towards the outer edge of the domain, significantly increasing the run time of our simulations. For the same reason, the transition region of magnetic field polarity within the equatorial plane of the domain, the heliospheric current sheet (HCS), is not additionally refined. We also note that the assumption of a reflection coefficient dominated by outwardly propagating waves ($\mathcal{R} < 1$) is equivalent to the assertion of a cross helicity not close to zero. While this is applicable for the inner heliosphere represented by our simulation domain, the cross helicity decreases systematically with heliocentric distance \citep[e.g.][]{roberts1987, adhikari2020, chen2020, chhiber2021}, which should be taken into account when extending the simulation domain of NIRwave (see also \citealt{cranmer2010}). 

Given the restrictions and simplifications constraining our simulations, we still find a good representation of the solar wind parameters in our model when compared to previous investigations and observations. The magnetic field settles into a structure dominated by the radial component $B_\mathrm{r}$ at distances beyond  $\sim$\SI{2}{\solarradius}, and almost all field regions are open within the domain. At the inner boundary, field lines above approximately \SI{33}{\degr} heliolatitude ($\theta = \SI{57}{\degr}$) are stretched out towards the outer boundary of the simulation domain and form open field regions, as also visible in Fig.~\ref{fig:3d-solarwind} (right panel). Within the larger scale of the whole simulation domain, all field regions with the exception of the HCS region centred at the equatorial plane are open. As the simulation grid is not additionally refined at this specific region, its size flares out towards the outer boundary of the simulation domain. 

We compare the radial magnetic field strength $\vert B_\mathrm{r} \vert$ resulting from our simulation with measurements taken by PSP. Specifically, \citet{telloni2021, telloni2022} give values for $\vert B_\mathrm{r} \vert$ during encounter phase 7, which these authors measured with the fluxgate magnetometer of the FIELDS suite of instruments \citep{bale2016} during two time intervals corresponding to heliocentric distances of \SI{21.4}{\solarradius} and \SI{23.6}{\solarradius}. They note values of $\vert B_\mathrm{r} \vert$ of \SI{2.38d-3}{\gauss} and \SI{1.44d-3}{\gauss} for these distances, respectively. We choose the absolute value of $B_\mathrm{r}$ as the spacecraft crosses the HCS during this encounter period (visible in Fig.~5 of \citealt{telloni2022}). We compare these measurements to our simulation results, which show values of approximately \SI{1.1e-3}{\gauss} and \SI{0.9d-3}{\gauss} at the same distance values. These are in agreement with the PSP measurements within a factor of two. Figure~\ref{fig:br_comparison} shows the profiles of $\vert B_\mathrm{r} \vert$ for the polar angle $\theta$ between the positive z-axis and the equator of the domain, also illustrating the extended size of our polarity transition region where the value of $\vert B_\mathrm{r} \vert$ approaches zero. As our model reasonably reproduces the observed solar wind parameters based on a simple dipole field assumption, it can be adapted to investigate the stellar winds of a larger sample of solar-like stars for which the dipole field component and other field components  are known, such as quadrupole and octopole modes (see \citealt{see2019}, and references therein).

\subsection{Comparison to polytropic model}\label{ssec:nirwave_polytrope}

To motivate the modifications we apply to NIRVANA in an effort to include an explicit WTD heating mechanism, we also compared the results from our solar wind model to a NIRVANA simulation conducted with a polytropic equation of state. As described in Sect.~\ref{ssec:wtd_source}, the code does not natively support the inclusion of a complex heating-source term representing a WTD mechanism. We therefore generated a magnetised polytropic solar wind model, using a constant polytropic index of $\Gamma = 1.1$ and a base coronal temperature of $T_0 = \SI{2}{\mega\kelvin}$ (see \citealt{johnstone2015a, vidotto2014b, vidotto2021} and references therein). We provide a more detailed description of this in Appendix~\ref{app:comparison_poly}. 

Figure~\ref{fig:nirwave_poly_comparison} illustrates a comparison between the results of our WTD wind simulation and the polytropic case through radial profiles for the major parameters of radial velocity ($v_\mathrm{r}$), number density ($n_\mathrm{p}$), and temperature ($T$) split into equatorial and polar radial profiles associated with the slow and fast wind, respectively. The radial velocity and density profiles for both the WTD and polytropic case follow largely similar trends. Compared to the WTD case, the fast wind of the polytropic model reaches a slower terminal velocity, and the acceleration of the polytropic slow wind is more sustained. This leads to a smaller difference in radial velocity between the slow and fast wind regimes in the polytropic case, specifically overestimating the slow-wind velocity seen in the empirical constraints to a larger degree than in the WTD case. For the density profiles, both solutions follow largely the same radial trend. However, the number density of the polytropic fast wind regime is approximately a factor of 2 larger than for the WTD case. The temperature profiles show a more significant difference between the two models: in the polytropic case, the radial behaviour of the temperature follows the density through Eq.~(\ref{eq:polytrope_rhoT}) and shows comparable values in the fast- and slow-wind regimes, while the WTD model displays a more pronounced difference between these two regimes.

In summary, the NIRVANA simulation results produced by a magnetised polytropic wind model are similar to the solar wind model we generate with NIRwave. As shown in Fig.~\ref{fig:nirwave_poly_comparison}, the radial profiles produced by these two approaches are largely in agreement given similar initial conditions, where the separation of parameter values between the slow- and fast wind regimes is less pronounced in the polytropic case. However, we argue that the implementation of a WTD heating mechanism in the solar wind model provides an improvement over the polytropic approach by introducing a physically motivated contribution to the energy equation in the form of wave dissipation rather than circumventing the question of coronal heating through the polytropic equation of state. Figure~\ref{fig:nirwave_poly_comparison} also illustrates that the results achieved with NIRwave are in better agreement with the empirical constraints derived from PSP observations than those found with the polytropic wind model produced with NIRVANA.

\begin{figure}
        \centering
        \includegraphics[width=\hsize]{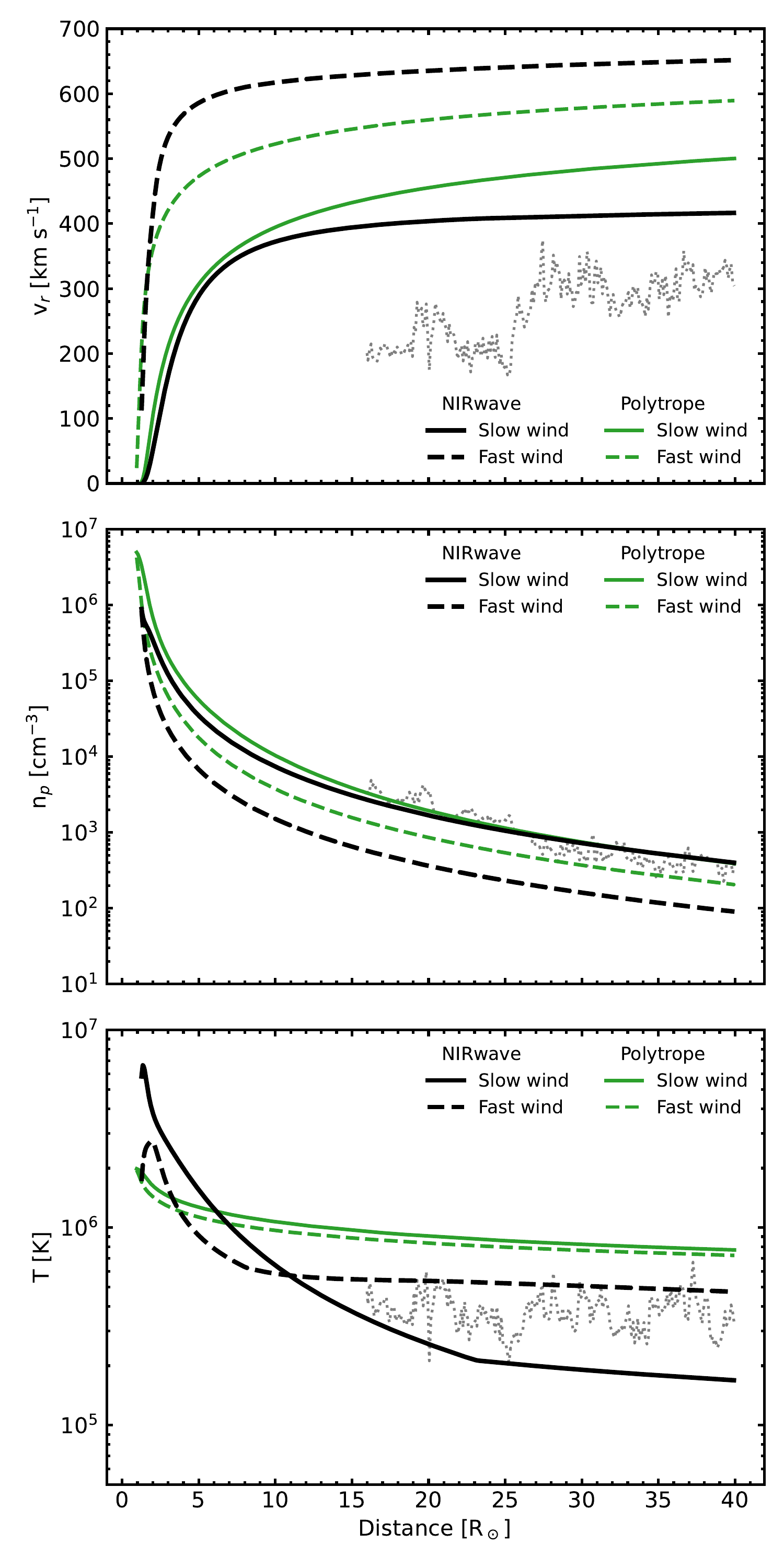}
        \caption{Comparison between the solar wind model generated with NIRwave (black lines) and a polytropic model using NIRVANA (green lines). The panels show the radial wind velocity ($v_\mathrm{r}$, top panel), number density ($n_\mathrm{p}$, middle panel), and temperature ($T$, bottom panel), each with respect to radial distance in \si{\solarradius} from the domain centre. The solid lines represent equatorial radial parameter profiles associated with the slow wind regime, and dashed lines represent polar radial parameter profiles associated with the fast wind. The grey dotted lines in each panel represent the observational data from PSP described in Sect.~\protect\ref{sec:observ_constraints}.}
        \label{fig:nirwave_poly_comparison}
\end{figure}

\section{Conclusions}\label{sec:conclusion}
We carried out solar wind simulations, using the general 3D MHD code NIRVANA to reconstruct the solar wind structure within \SI{40}{\solarradius}. To account for an explicit heating mechanism as a driving force of the solar wind, we modified NIRVANA to include an established wave-turbulence-driven heating routine as an energy-source term for the solar wind acceleration. We approximated the solar magnetic field with a dipole, and used observations taken by PSP to validate our simulation results.

The simulation results achieved in this work are in reasonable agreement with the measured wind properties. Our model reproduces the characteristic bimodal structure of the solar wind and accounts for the radial behaviour of the solar wind velocity ($v_\mathrm{r}$), proton number density ($n_\mathrm{p}$), and temperature ($T$). The slow and fast wind regimes of our steady-state solution are characterised by terminal velocities of \SI{410}{\kilo\meter\per\second} and \SI{650}{\kilo\meter\per\second}. We find a mass-loss rate of \SI{2.6e-14}{\solarmass\per\year} in agreement with previously established results, and ram-pressure values that follow the constraints we derived from PSP measurement data. We also find radial magnetic-field-strength values in good agreement with evaluations of measurements taken by PSP. The Alfvén critical surface within our simulation domain is smaller by a factor of approximately 3 than predicted by previous models and observed by PSP.

While our results globally agree with PSP observations, slight differences are seen in the absolute values. The precision of our simulation results is constrained through limitations in the grid setup and simplifying assumptions for the conditions of the solar wind. The simulation grid is uniformly decomposed, which restricts the resolution of regions with small radial values and close to the equator of the domain. We introduced the solar magnetic field as a pure dipole, which overlooks the complexity of the solar photospheric and chromospheric magnetic field structure observable through solar magnetograms. Including such magnetograms as boundary conditions for the magnetic field provides a way to potentially increase the complexity of our model in the future. Additionally, the modular implementation of user-defined heat-source terms into NIRVANA can also facilitate the inclusion of additional or other heating terms as a driving force of the solar wind, potentially allowing comparability between the achieved results.

Despite the simplified conditions we assumed for our model, we achieve results for the distribution of solar wind parameters that are comparable to those of other, more complex wind models. Developing an accurate model of the solar wind constrained by {in situ} measurements is fundamental for the investigation of stellar-wind parameters in solar-like stars. While not part of this work, a larger sample of low-mass, Sun-like stars with known dipole field components could be explored in future work through variation of the input parameters that characterise these stars.

\begin{acknowledgements}
        
      We thank the anonymous referee for their valuable suggestions and comments. We also thank M. Stevens and R. Livi for valuable comments on the usage of SPC and SPAN-I data. S.B.S acknowledges funding by the Austrian Science Fund (FWF) through the Lise-Meitner grant M~2829-N. Parker Solar Probe was designed, built, and is now operated by the Johns Hopkins Applied Physics Laboratory as part of NASA’s Living with a Star (LWS) program (contract NNN06AA01C). Support from the LWS management and technical team has played a critical role in the success of the Parker Solar Probe mission. Thanks to the Solar Wind Electrons, Alphas, and Protons (SWEAP) team for providing data (PI: Justin Kasper, BWX Technologies). The results of this work were partially achieved at the Vienna Scientific Cluster (VSC).
      
\end{acknowledgements}

\bibliography{references.bib}
\bibliographystyle{aa}

\begin{onecolumn}
\begin{appendix}

    \begin{figure*}[!t]
        \section{PSP observational data}
        \centering
        \includegraphics[width=\textwidth]{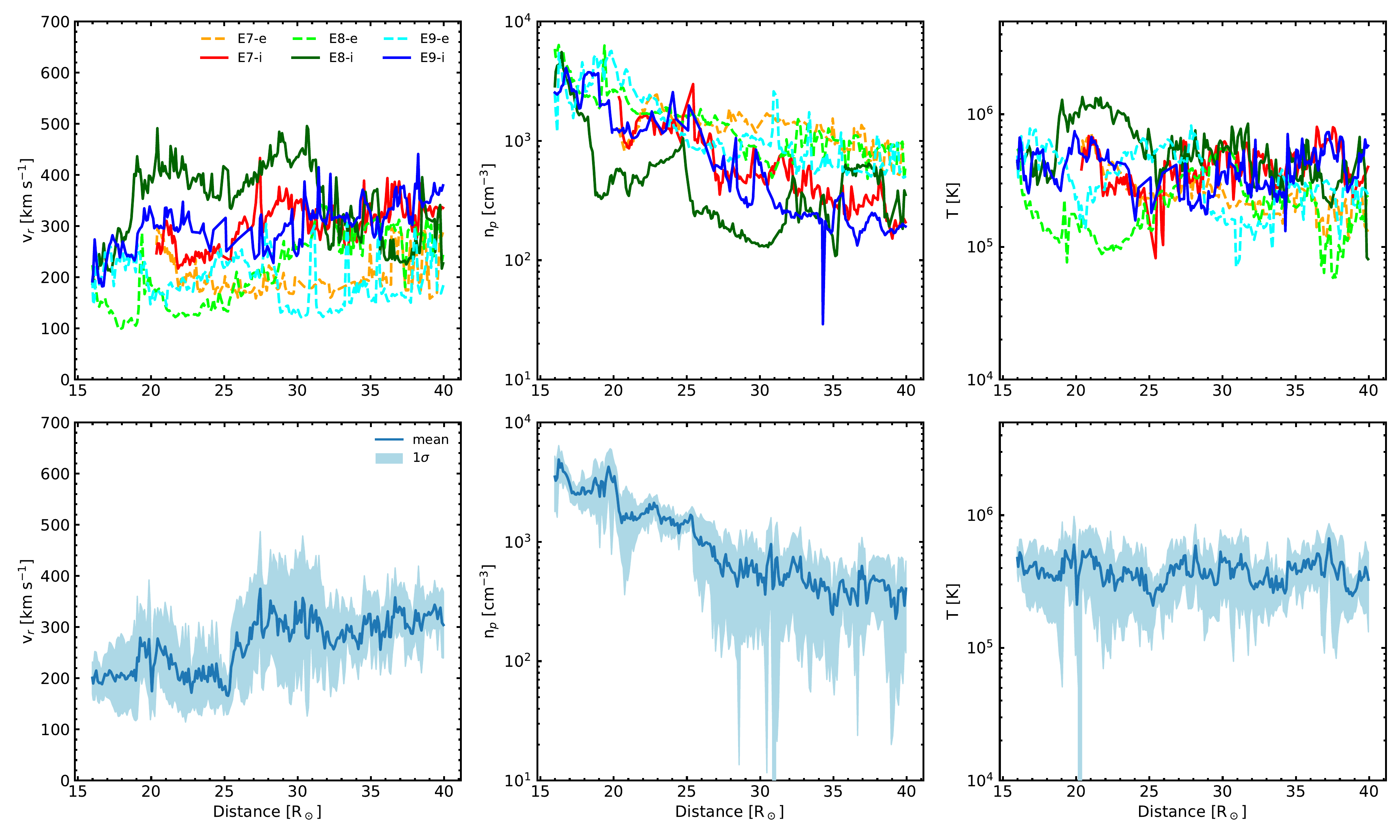}
        \caption{Evaluated measurement data from SPC and SPAN-I for the encounter periods 7, 8, and 9. Top: Radial velocity $v_\mathrm{r}$ (left panel), proton number density $n_\mathrm{p}$ (middle panel), and temperature $T$ (right panel) measurements. All three plots are split into ingress (solid lines) and egress (dashed lines) periods of the spacecraft. The observed parameters are display in a distance-binned form with a bin size of $\Delta r = \SI{1e-1}{\solarradius}$.
        Bottom: Same arrangement as above, but showing the evaluated mean values (solid lines) and 1$\sigma$ deviation (filled-in area).}
        \label{fig_app:spc_ingress_egress_obs}
    \end{figure*}

    
    \begin{figure}[!t]
    \section{Number of PSP data points}
        \centering
        \includegraphics[width=0.5\textwidth]{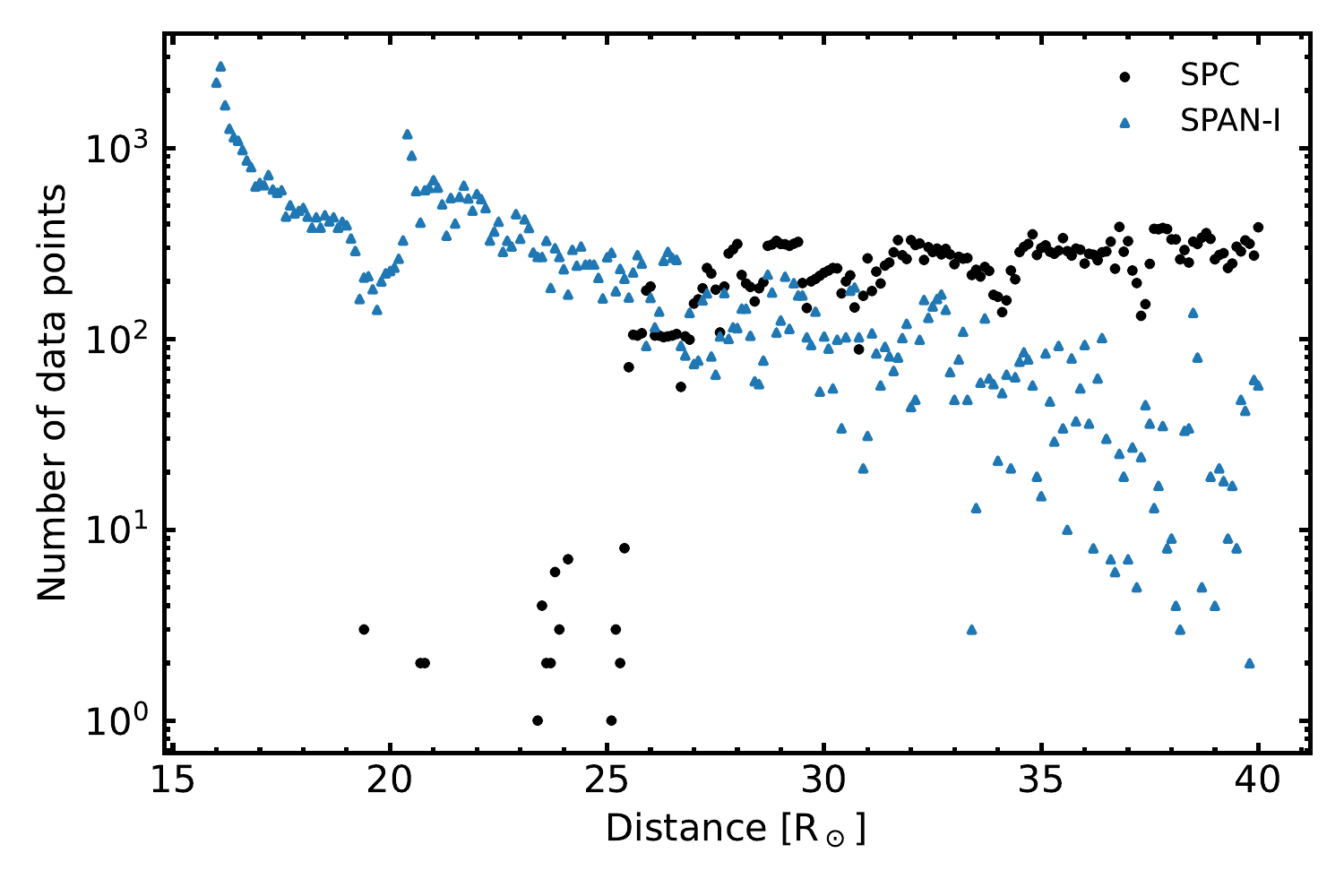}
        \caption{Number of ten-second-averaged measurements within each distance bin $(\Delta r = \SI{0.1}{\solarradius})$ after data reduction of the available SPC (black dots) and SPAN-I (blue triangles) observations for encounter 7, 8, and 9.}
        \label{fig:data_per_bin}
    \end{figure}
    
    
    \begin{figure*}[!t]
        \section{Encounter 8 Ingress}
        \centering
        \includegraphics[width=\textwidth]{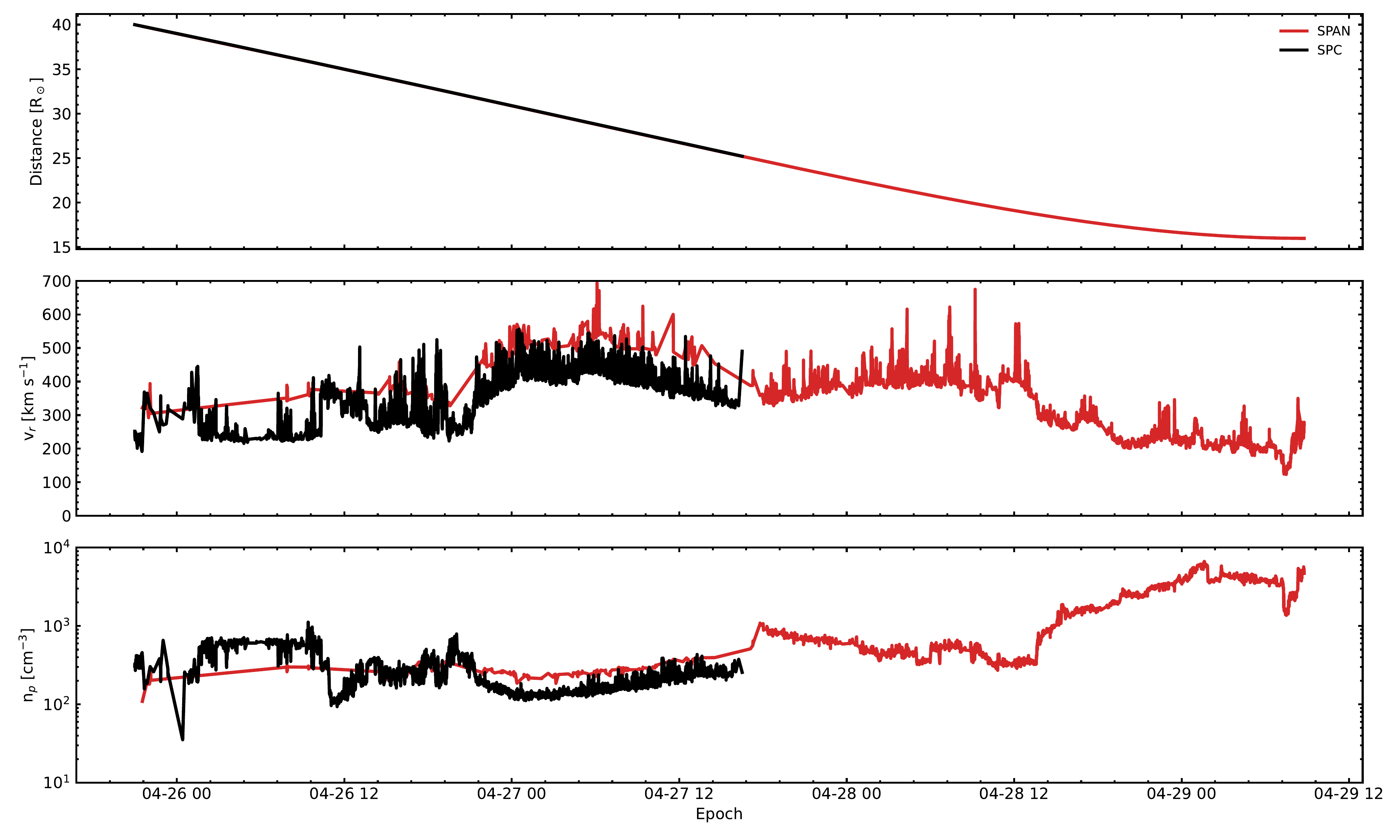}
        \caption{SPC (black line) and SPAN-I (red line) measurements for the ingress phase of the designated encounter period 8 (restricted to heliocentric distances $< \SI{40}{\solarradius}$). The panels show heliocentric distance (top panel), radial velocity component (middle panel), and number density (bottom panel) with respect to time, starting at approximately 26 April 2021 and ending at approximately 29 April 2021. The displayed measurements indicate the complementary nature of SPC and SPAN-I, transitioning at approximately \SI{25}{\solarradius}.}
        \label{fig_app:psp_8i_obs}
    \end{figure*}


    \begin{figure}[!t]
    \section{Radial magnetic field strength}
        \centering
        \includegraphics[width=0.5\textwidth]{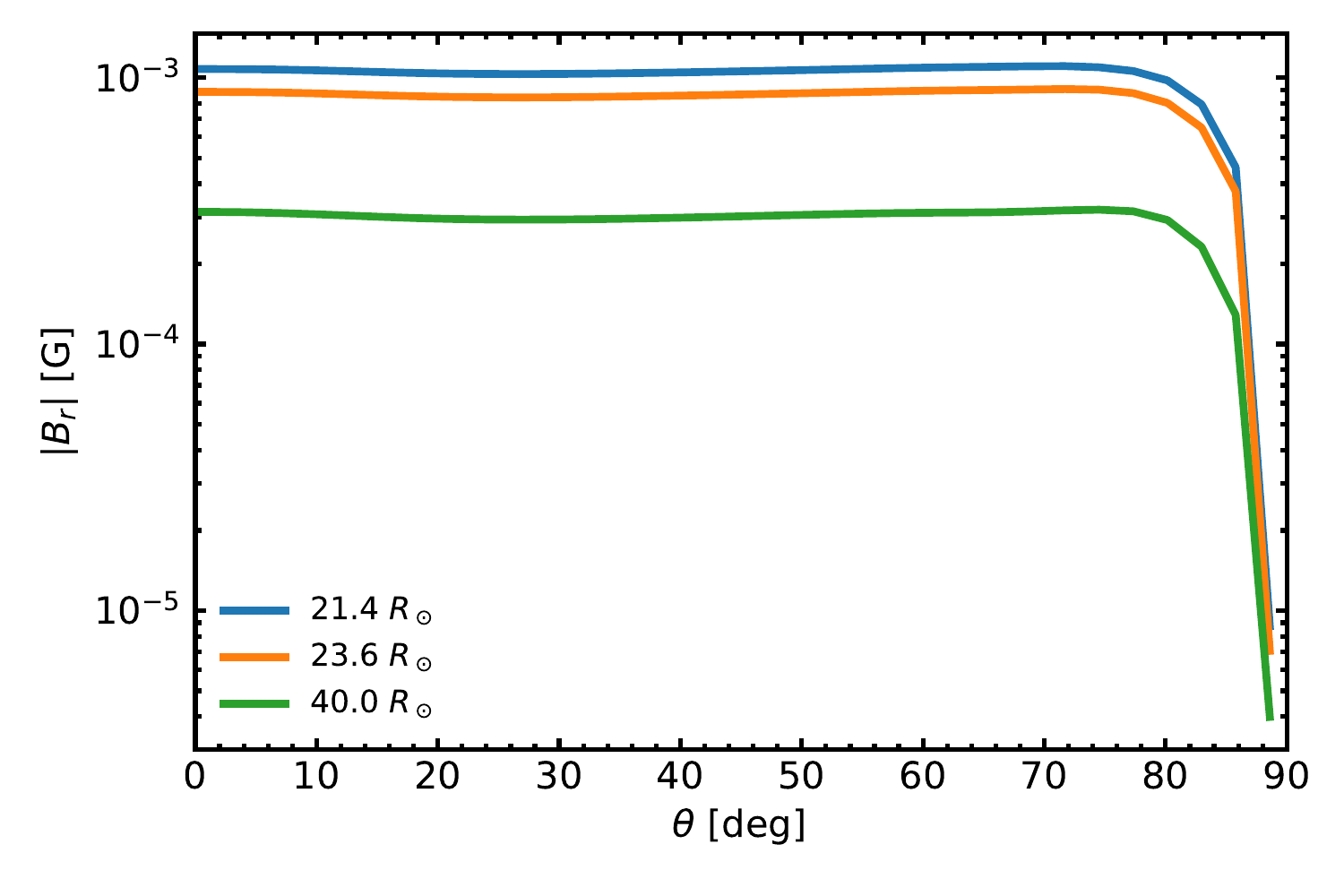}
        \caption{Radial magnetic field strength $\vert B_\mathrm{r} \vert$ with respect to polar angle $\theta$ of our simulation result for radial distances of \SI{21.4}{\solarradius} (blue line), \SI{23.6}{\solarradius} (orange line), and \SI{40}{\solarradius} (green line).}
        \label{fig:br_comparison}
    \end{figure}

    \FloatBarrier
    \section{Polytropic wind model}\label{app:comparison_poly}
    
    We generate a magnetised polytropic wind solution with NIRVANA using a polytropic equation of state, meaning the density and thermal pressure in this case are related through a power law defined by the polytropic index $\Gamma$ and the polytropic constant $K$:
    \begin{equation}
        p = K \rho^\Gamma.
    \end{equation}
    
    The value of $K$ is defined by the density, $\rho$, and temperature, $T$, of the solar wind,
    \begin{equation}\label{eq:polytrope_rhoT}
        K = \frac{k_\mathrm{B} \rho^{1-\Gamma}}{\bar{\mu} m_\mathrm{p}} T,
    \end{equation}
    where $k_\mathrm{B}$ is the Boltzmann constant, $\bar{\mu}$ the mean molecular weight, and $m_\mathrm{p}$ the proton mass.
    
    For this polytropic wind simulation, we use the same setup as described in Sect.~\ref{sec:model_desc} concerning boundary and initial conditions (where applicable). Additionally, we choose a constant polytropic index of $\Gamma = 1.1$, coinciding with a value appropriate for the regions close to the Sun and determine the value of the polytropic constant by choosing a temperature of $T_0 = \SI{2}{\mega\kelvin}$ to represent coronal values at the inner boundary (see \citealt{johnstone2015a, vidotto2014b, vidotto2021} and references therein). The steady-state solution of this polytropic simulation is illustrated in Fig.~\ref{fig:poly_structure}, showing the meridional and equatorial distribution of $v_\mathrm{r}$ together with the global magnetic field structure (left panel). Comparing this to the left panel of Fig.~\ref{fig:3d-solarwind} shows that in both cases, the global magnetic field structure is predominantly radial. The difference in results between the two cases becomes more apparent in the distribution of $v_\mathrm{r}$, making up the slow and  fast wind regimes. Comparing the right panel of Fig.~\ref{fig:poly_structure} with Fig.~\ref{fig:2d_solarwind} illustrates that the fast wind regime of the polytropic simulation evenly extends towards small heliolatitudes, where it transitions into the slow wind regime through a steep velocity gradient. This stands in contrast to the bimodal distribution of $v_\mathrm{r}$ from the wave-turbulence-driven model (visible in Fig.~\ref{fig:2d_solarwind}), which shows a more gradual transition between the fast and slow wind regimes. This difference in the radial-velocity distribution is also reflected in the shape of the Alfvén surface, which shows a more spherical contour following the extended distribution of the fast wind. The scale of the Alfvén surface is similar to the result from NIRwave, where we refer to Sect.~\ref{ssec:nirwave_obs} for an explanation. A comparison of the radial profiles for the fast and slow wind regimes of both the WTD and polytropic solution are shown in Fig.~\ref{fig:nirwave_poly_comparison} and described in Sect.~\ref{ssec:nirwave_polytrope}.

    \begin{figure*}[!h]
        \centering
        \includegraphics[width=\textwidth]{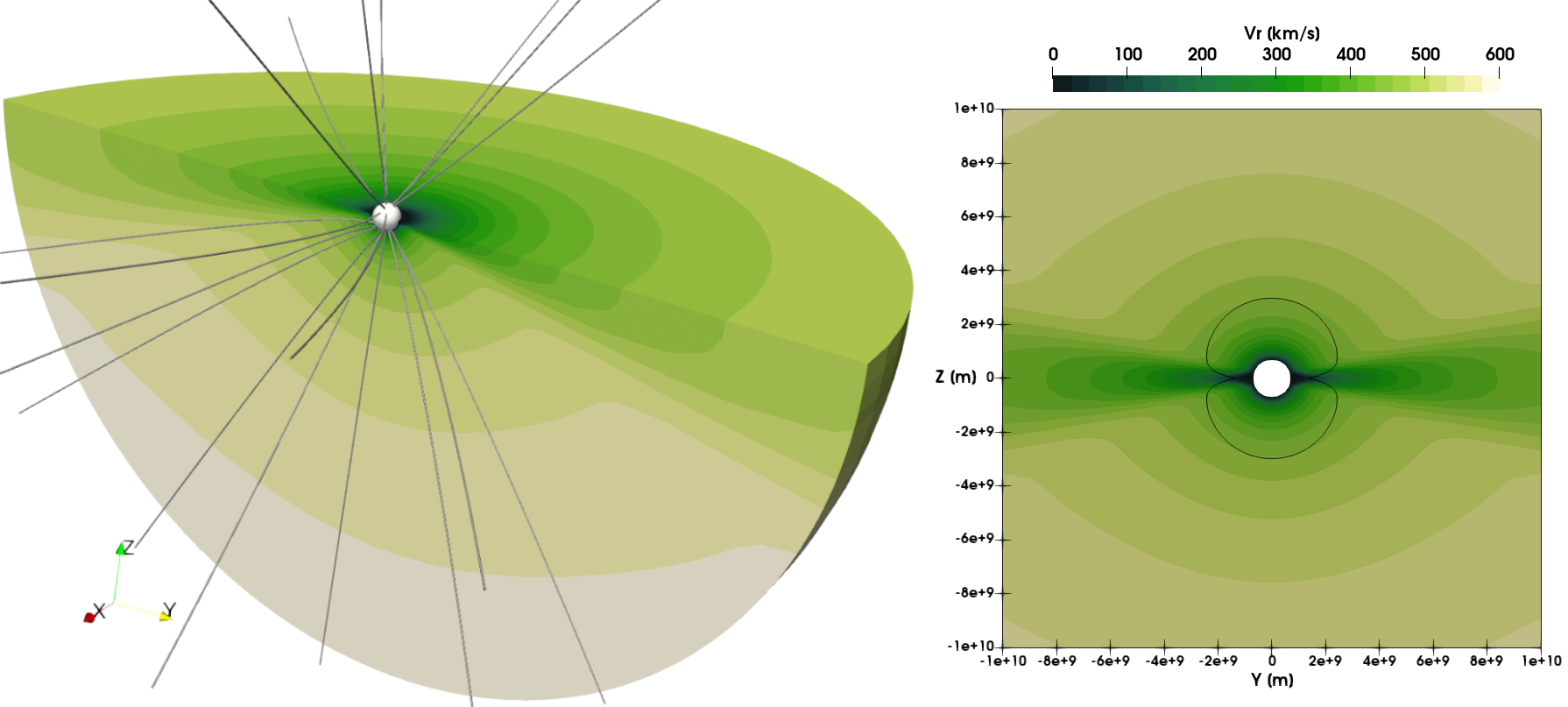}
        \caption{Radial-velocity structure of our polytropic solar wind simulation. Left: Three-dimensional radial-velocity structure (colour map) and magnetic field lines (grey lines). Right: Meridional slice of the simulation domain (X = 0) extending to $\pm\SI{15}{\solarradius}$ in both Y and Z directions. The solid black outline denotes the location of the Alfvén surface. The colour map displaying radial-velocity values is shared for the left and right figure.}
        \label{fig:poly_structure}
    \end{figure*}


\end{appendix}
\end{onecolumn}

\end{document}